# Routing Valley Excitons in a Monolayer MoS$_2$ with a Metasurface

Liuyang Sun[1], Chun-Yuan Wang[1,2], Alexandr Krasnok[3], Junho Choi[1], Jinwei Shi[4], Juan Sebastian Gomez-Diaz[3,5], Andre' Zepeda[1], Shangjr Gwo[2,6], Chih-Kang Shih[1], Andrea Alù[3,7*], Xiaoqin Li[1*]

[1] *Department of Physics, Complex Quantum Systems, and Texas Materials Institutes, University of Texas at Austin, Austin, TX 78712, USA*

[2] *Department of Physics, National Tsing-Hua University, Hsinchu 30013, Taiwan*

[3] *Department of Electrical and Computer Engineering, The University of Texas at Austin, Austin, Texas 78712, USA*

[4] *Department of Physics and Applied Optics Beijing Area Major Laboratory, Beijing Normal University, Beijing 100875, China*

[5] *Department of Electrical and Computer Engineering, University of California, Davis, Davis, CA 95616, USA*

[6] *National Synchrotron Radiation Research Center, Hsinchu 30076, Taiwan*

[7] *CUNY Advanced Science Research Center, New York, NY 10031, USA*

\* Andrea Alù, alu@mail.utexas.edu

\* Xiaoqin Li, elaineli@physics.utexas.edu

**Abstract**

Excitons in monolayer transition metal dichalcogenides (TMDs) are formed at *K* and *K′* points at the boundary of the Brillouin zone. They acquire a valley degree of freedom, which may be used as a complementary platform for information transport and processing[1-3]. In a different context, metasurfaces consisting of engineered arrays of polarizable inclusions have enabled the manipulation of light in unprecedented ways, and found applications in imaging[4-8], optical information processing[9-11], and cloaking[12-15]. Here, we demonstrate that, by coupling a $MoS_2$ monolayer to a suitably designed metasurface consisting of asymmetric grooves, valley polarized excitons can be sorted and spatially separated even at room temperature. Emission from valley excitons is also separated in *K*-space, i.e., photons with opposite helicity are emitted to different directions. Our work demonstrates that metasurfaces can facilitate valley transport and establish an interface between valleytronic and photonic devices, thus addressing outstanding challenges in the nascent field of valleytronics.

The extrema of the energy-momentum dispersion are called valleys. Typically, it is not possible to control individual valleys because they are not strongly coupled to any external field or force. An exception is found in monolayer TMDs, where the broken inversion symmetry combined with time-reversal symmetry lead to opposite spins at the $K$ and $K'$ valleys, effectively locking the spin and valley degrees of freedom (DoF) [16-18], as illustrated in Fig. 1a. Thus, optical manipulation of the valley DoF can be realized via exciton resonances based on the valley contrasting optical selection rules. This unique opportunity to address valley index make it possible to explore this binary quantum degree of freedom as an alternative information carrier [1-3] that may complement both classical and quantum computing schemes based on charge and spin.

As a prerequisite to building valleytronic devices, a number of previous experiments have explored strategies to separate valley-polarized free carriers or excitons. For example, free carrier valley Hall effect in monolayer $MoS_2$ was first demonstrated using electric readout method[19]. There, the Berry curvature of the energy band acts as a momentum-dependent magnetic field, leading to transverse motions of valley polarized carriers in the presence of an in-plane electric field. Similarly, exciton valley Hall effect was recently reported for a monolayer $MoS_2$ where the exciton motion is driven by a temperature gradient [20]. However, other factors such as uncontrolled strain can also lead to the separation of valley-polarized excitons. These previously reported carrier and exciton valley Hall effects were observed at low temperature. Since the binding energy of excitons is very large in TMDs ( a few hundred meVs) [21-25], it is therefore both feasible and highly desirable to design a photonic device to manipulate valley excitons at room temperature.

In this paper, we demonstrate that valley excitons in a monolayer MoS$_2$ can be spatially separated at room temperature when placed on top of a suitably designed metasurface (as illustrated in Fig.1b). The spatial separation of valley excitons of opposite chirality is enabled by coupling to surface plasmon polaritons (SPPs) propagating along asymmetrically shaped grooves [26]. This method to separate valley index is rather general, and it can be applied to a wide range of layered materials. No exciton valley polarization is required, as is the case for MoS$_2$ at room temperature. Furthermore, photons with definite chirality emitted by valley excitons are separated in momentum space, enabling far-field optical detection of valley excitons, and therefore serving as an interface between valleytronics and photonic devices.

Unidirectional launching of SPPs based on the photonic spin Hall effect has been previously demonstrated at a metal-dielectric interface and in metamaterials[27-30]. Specifically, opposite circularly polarized dipoles have been shown to excite SPPs propagating towards opposite directions. All these demonstrations so far, however, have focused on circularly polarized dipoles with an out-of-plane component, which couples with transverse-magnetic (TM) polarized SPPs propagating along the plasmonic interfaces. The K and K' valley excitons in a monolayer MoS$_2$ can be modeled as in-plane circularly polarized dipole ($E_x \pm iE_y$) oscillating with opposite helicity, as illustrated in Fig. 1a. Their in-plane oriented dipoles, however, cannot asymmetrically excite conventional SPPs, since they do not engage the required out-of-plane chirality. Thus, it is necessary to rely on specifically designed metasurfaces that can break chiral symmetry within the plane, and that can enhance the coupling efficiency additionally [31].

We designed a metasurface consisting of asymmetrically shaped grooves arranged in a subwavelength period, as shown in Fig. 2. Each groove supports SPP propagation along its side walls. In the case of an array of symmetric grooves (Fig. 2a), the mirror symmetry along the y-z plane (dashed black curve) leads to chirality independent propagation, i.e., an in-plane dipole with left- or right-handed chirality couples to both sidewalls equally. However, if the mirror symmetry is intentionally broken by tilting one side wall at an angle (sketched in Fig. 2b), the SPP excitation induced by valley excitons is different for vertical and tilted walls, respectively, resulting in net unidirectional propagation of SPPs with opposite circularly polarized electric fields *in the plane* of the metasurface (see SI for details). This crucial concept extends the photonic spin-Hall effect to planar metasurfaces with only in-plane electric field and enables chirality dependent coupling between TMDs and metasurfaces. Figs. 2c-e compare the spatial distribution of electric field intensity for the symmetric groove metasurface (both in-plane circular polarizations produce the same response, Fig. 2c), and for asymmetric grooves, for which chirality dependent SPP propagation is observed (Figs. 2d-e) excited by a single, circularly polarized in-plane dipole. The groove array with subwavelength period enables better spectral overlap between the SPP resonance and the $MoS_2$ excitons, boosting the SPP-exciton coupling efficiency. More importantly, the periodic array allows separation of valley excitons independent of the exact position of the excitation laser on the plane of the metasurface, more robust than those schemes relying on a single plasmonic nanostructure [32, 33] (e.g. a nanowire).

The metasurface was fabricated on high-quality Ag plates grown by a chemical synthesis method developed by us[34] using focused ion beam (FIB) milling. The SEM image from a region of the metasurface is shown in Fig. 3a. A single asymmetric groove is fabricated in two FIB steps, leading to two different depths ($d_1$ = 150 nm and $d_2$ = 90 nm) and a total width ($w$) of 90 nm. The depth, width and period (200 nm) determine the spectral position of the SPP resonance. After FIB milling, a conformal $Al_2O_3$ layer (~5 nm thick) was deposited using atomic layer deposition to protect the Ag metasurface structures and to prevent photoluminescence (PL) quenching of the MoS$_2$ monolayer.

The monolayer MoS$_2$ was mechanically exfoliated and transferred onto the metasurface. Part of the monolayer MoS$_2$ covers the groove array, while part of the monolayer is on top of the unstructured Ag film. Measurements are taken from both regions for comparison. We refer to measurements taken on the smooth Ag surface as the control sample. We first take PL spectra using a circularly polarized excitation laser centered at 532 nm. It was shown in previous experiments that a MoS$_2$ monolayer excited by a $\sigma_+$ polarized laser preferentially emits photons with $\sigma_+$ helicity at cryogenic temperature[16-18, 35]. The degree of valley polarization is quantified as $\rho_{(\omega)} = \frac{I_+(\omega) - I_-(\omega)}{I_+(\omega) + I_-(\omega)}$, where $I_+$ ($I_-$) refers to the PL intensity with $\sigma_+$ ($\sigma_-$) polarization at the peak of the exciton resonance.

Exciton valley polarization strongly depends on the excitation laser wavelength and temperature[16-18, 35]. With an excitation laser at 532 nm and at room temperature, we did not observe any exciton valley polarization, consistent with previous experiments (data shown in the supplementary). By placing MoS$_2$ on top of a metasurface, not only valley excitons are separated in real space, the emitted photons with different helicity are also separated in momentum space, as we demonstrate below. We use $r$ and $k$ to represent real space and photon momentum space, respectively.

To observe the spatial separation between valley excitons (as illustrated in Fig. 3b), we first choose a linearly polarized excitation laser to populate two valleys equally. The linear polarization is perpendicular to the groove to enhance the local electric field along the edges of the grooves, leading to increased absorption by the MoS$_2$ monolayer. The spatial profile of the PL from MoS$_2$ monolayer on metasurface, monolayer on a smooth Ag surface, and the excitation laser itself are all captured by a CCD camera following a spectrometer. We select the $\sigma_+/\sigma_-$ PL by placing a series of polarizers and waveplates in the detection path. The $\sigma_+$ and $\sigma_-$ PL images are taken separately. The difference between the two spectra is then divided by their sum to obtain a spatial image of valley polarization, $\rho(r)$. The measured $\rho(r)$ from the MoS$_2$ monolayer placed on the metasurface is shown in Fig. 3c, in which a clear spatial separation of valley excitons is observed. In contrast, no detectable valley polarization $\rho(r)$ was observed from the monolayer MoS$_2$ on the unstructured silver film (Fig. 3d).

To quantify the spatial separation, we plot the differential intensity profile $\Delta I(y) = I_+(y) - I_-(y)$ along a white dashed line parallel to the groove direction in Fig. 3e. The sign of $\Delta I(y)$ changes across the center of the laser spot at $y = 0\ \mu m$. The profile can be reproduced well by subtracting two Gaussian functions centered at $y = \pm 0.7 \mu m$ respectively as a dashed black curve in Fig. 3e. The separation of the peaks reaches $\sim 1.4\ \mu m$. Figure 3f is the numerical simulation of the spatial valley polarization along an asymmetrically shaped groove.

In addition to valley exciton spatial separation, we also expect that the emitted chiral photons are separated in momentum space, a desirable feature for interfacing valleytronic with photonic devices. Because of the unidirectional launching, and the specific metasurface design that allows radiation leakage of the SPPs, polarized waves with opposite helicity will out couple to different directions, leading to separation of leaked and scattered energy in momentum space. As illustrated in Fig. 4a, photons with opposite helicity are emitted to different directions. The simulated far-field emission pattern is shown in Fig. 4b. The $\sigma_+$ and $\sigma_-$ polarized photons preferentially emit toward the upper and lower hemisphere as indicated in the polar-plot. We define $\rho(k_\parallel)$ to describe the chirality dependent emission pattern quantitatively. Here, $k_{||} = k_0 \cdot sin\theta$ is the momentum component of light in y-z plane, and $k_0 = 2\pi / \lambda$ is the wavevector of light at wavelength $\lambda$ in free space. Experimentally, we obtained $\rho(k_{||})$ by the $k$-space imaging technique. The setup is sketched in Fig. 4c and explained in detail in the method.

Similarly, the $\rho(k_{||})$ is obtained by taking the $k$-space images of $\sigma_+$ and $\sigma_-$ emissions separately. The ratio between the difference and the sum of the two images from monolayer $MoS_2$ on metasurface is plotted in Fig. 4d. The sign of $\rho(k_{||})$ reverses at the opposite sides in the k space, demonstrating that the metasurface leads to directional emission of the chiral photons. In contrast, $\rho(k_{||})$ is negligible across the whole k space for photons emitted by a monolayer $MoS_2$ placed on a smooth Ag surface as show in Fig. 4e.

In summary, we demonstrated that an achiral metasurface may be used to route valley polarized excitons. Our method is based on a specifically tailored metasurface consisting of subwavelength arrays of asymmetric grooves which can separate circular polarizations in the plane of the metasurface. The metasurface design is generally applicable to atomically thin materials that do not exhibit any valley polarization at room temperature. The degree of valley polarization and its propagation distance can be further improved by optimizing the metasurface design (e.g., via chiral launching sites) and fabrication. The combination of metasurfaces as passive components and 2D materials as active components enables conceptually novel hybrid photonic devices in a compact geometry. Such hybrid devices may be used to control exciton/spin/valley transport[36] in unprecedented ways and to engineer quantum emitter arrays[37].

**Acknowledgements** The spectroscopic experiments performed at UT-Austin are supported by NSF EFMA-1542747 and NSF DMR-1306878. J.C., C.S, X.L. are supported by NSF MRSEC program DMR-1720595. X. L. also gratefully acknowledges the support from the Welch Foundation via grant F-1662. A.A. acknowledges the Air Force Office of Scientific Research and the Welch Foundation with grant No. F-1802. S.G. and C-Y.W. acknowledge the support from the Ministry of Science and Technology (MOST) in Taiwan (MOST 105-2112-M-007-011-MY3). The collaboration between National Tsing-Hua University and The University of Texas at Austin is facilitated by the Global Networking Talent (NT3.0) Program, Ministry of Education in Taiwan.

**Author Contributions** L.S. led the optical experiments. C.W. synthesized Ag plates and fabricated the metasurface. A.K. performed the simulations. J.C. J.S., A.Z. assisted the experiments. L.S., C.W., A.K., X.L., and A.A. wrote the manuscript. X.L., A.A., C.S., and S.G. designed and supervised the project. All authors discussed the results and commented on the manuscript at all stages.

**Author Information** Supplementary information is available in the online version of the paper. Reprints and permission information is available online at www.nature.com/reprints. Correspondence and requests for materials should be addressed to X.L. (elaineli@physics.utexas.edu) and A.A (aalu@gc.cuny.edu).

**Competing Financial Interests** The authors declare no competing financial interests.

**Method:**

**Metasurface fabrication:**

The metasurface is created on top of a single crystalline Ag crystal synthesized in solutions. The single crystalline Ag crystals exhibit low loss because of its atomically smooth surface and absence of grain boundary. Optical properties of these Ag plates were carefully characterized. It has been shown that surface plasmon polariton can propagate beyond 100 μm on an unstructured Ag plate.

**Optical Measurements:** Polarization-resolved photoluminescence (PL) images were collected using a home-built micro-PL setup. A continuous-wave laser at 532 nm was used to excite the monolayer $MoS_2$. The polarization of the incident light was controlled by a linear polarizer (GTH5M, Thorlabs) combined with a half-wave plate (WPH10M-532, Thorlabs). The incident laser was focused onto the sample by a 100x objective lens (Mitutoyo Plan Apo) after reflected by a non-polarizing beam splitter (BS013 Thorlabs). The PL was collected by the same objective lens, transmitted through the beam splitter, a quarter-wave plate (05RP32), a linear polarizer (LPVIS050) and long pass filters, and entered spectrometer equipped with an CCD camera. In order to obtain *k*-space image, an additional lens and iris aperture were placed after the objective lens (see supplementary for sketch of the setup).

**Numerical Simulations**: FDTD simulations of the Ag-$MoS_2$ structure have been conducted by using CST Microwave Studio 2017. CST Microwave Studio is a full-wave 3D electromagnetic field solver based on finite-integral time domain solution technique. A nonuniform mesh was used to improve the accuracy near the Ag slab where the field concentration was significantly large and inhomogeneous. The measured permittivity data for single crystalline Ag are used [38]. The 2D $MoS_2$ has been modeled as 0.7 nm thick dielectric layer with experimentally obtained permittivity [39].

**Supplementary Information**

Reflection spectra from metasurface

PL spectra from MoS$_2$ monolayer on smooth Ag film

Image of PL, laser spot and optical microscopic image

Optical setup

Analytical Model

Figures

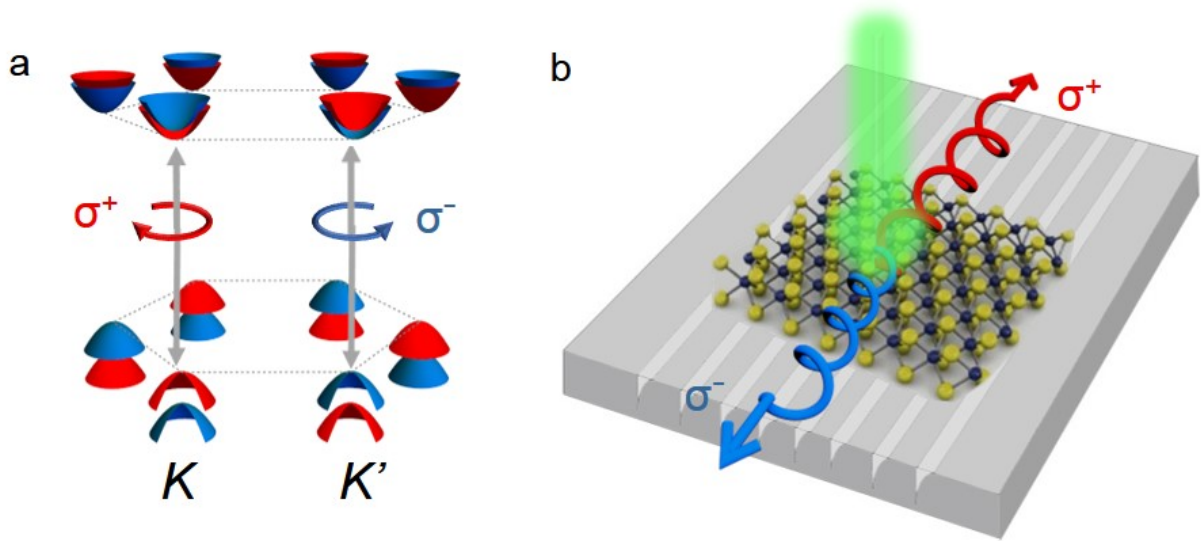

Fig.1 Schematics of the optically addressable valleys and spatial separation of valley excitons by a metasurface. a) Schematics of the band structure and optical selection rules of excitons in monolayer $MoS_2$. $\sigma^+$ and $\sigma^-$ polarized light preferentially couples to excitons in $K$ and $K'$ valleys, respectively. b) Illustration of valley excitons in a monolayer TMD controlled by a metasurface consisting of asymmetric grooves. Not only valley polarized excitons are spatially separated, photons with opposite helicity are also emitted to different directions, serving as a valley-photon interface mediated by excitons.

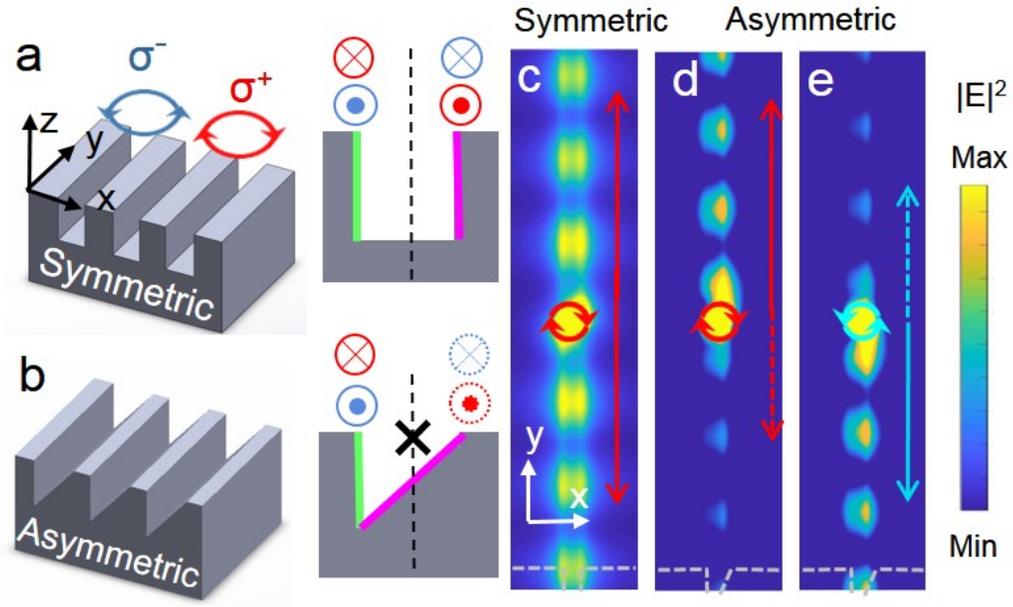

Fig. 2 Metasurface design principle. Illustration of the SPPs propagation excited by circularly polarized light in a) symmetric and b) asymmetric grooves. On the left wall, SPPs excited by a $\sigma_+$ ($\sigma_-$) dipole propagate into (out of) the page. On the right wall, the direction of propagation reverses. For the a) symmetric grating, no chirality dependent SPP propagation is observed due to the mirror symmetry indicated by the dashed black line. The mirror symmetry is lifted for b) an asymmetric groove resulting in chiral dependent SPP propagation. Simulated electric field intensity distribution induced by a $\sigma_-$ dipole on c) a symmetric and d) an asymmetric grating. e) Electric field intensity distribution induced by a $\sigma_+$ dipole on the asymmetric grating. Grey dashed lines in (c-e) indicate the position of the side walls.

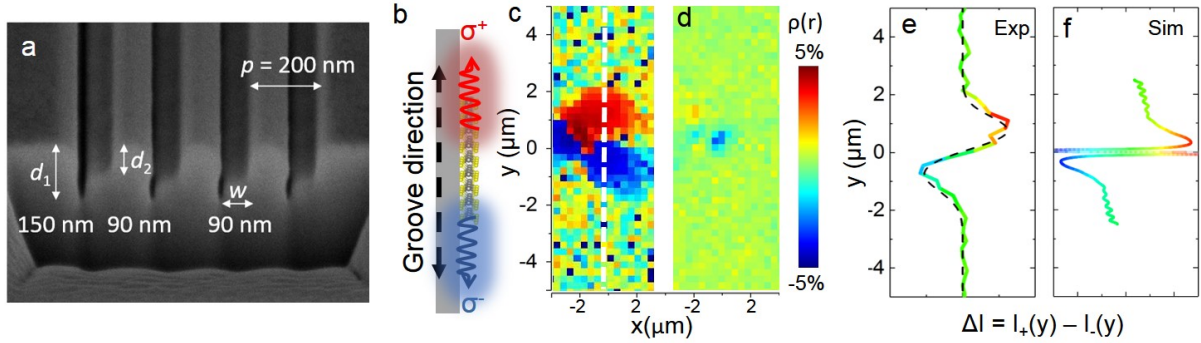

Fig. 3. Experimental observation of spatial separation of valley excitons by a metasurface consisting of asymmetric grooves. a) SEM image of the cross section of the asymmetric grooves. b) Illustration of valley exciton separation caused by the metasurface. Real space color plot of valley polarization $\rho(\boldsymbol{r})$ obtained from the c) MoS$_2$- metasurface and d) MoS$_2$-flat silver film. e) A line profile $\Delta I(y) = I_+(y) - I_-(y)$ along the white dash line in c). Black dash curve is a fitting by subtracting two Gaussian curves. f) A line profile from simulation shown in Fig. 2. The sharp peak in the middle is an unphysical divergence due to the simple point dipole model.

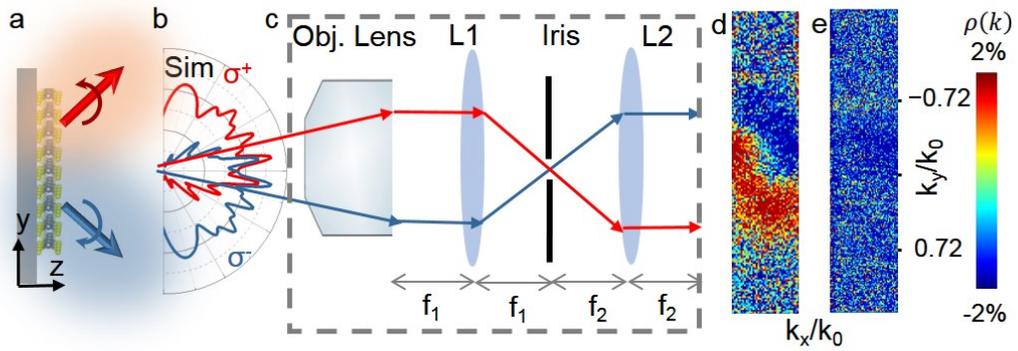

Fig. 4. Experimental observation of valley exciton emission separation in momentum space caused by the metasurface. a) Illustration of the helicity-dependent directional emission of the valley excitons. b) Numerically simulated far-field emission pattern from valley excitons with opposite helicity. c) Set-up for the k-space mapping of PL. Experimental $\rho(k_{\parallel})$ distribution in photon momentum space obtained from d) MoS$_2$-metasurface e) MoS$_2$-flat silver film. $k_x / k_0$ has the same ration with $k_y / k_0$.